\documentclass{article}

\usepackage{graphicx}  
\usepackage{amsmath}   
\usepackage{amssymb}   
\usepackage{bm}

\usepackage{relsize}  

\def\eq#1{{Eq.~(\ref{#1})}}

\def\zpl{{zero-point-length}}

\def\bk#1#2#3{{\langle #1|#2|#3\rangle}}  
\def\amp#1#2{\langle #1 | #2\rangle}      

  \title{Principle of Equivalence at Planck scales, QG in locally inertial frames and the zero-point-length of spacetime}
    
    \author{
  T. Padmanabhan\\
  {\small IUCAA, Pune University Campus,
  Ganeshkhind, Pune - 411 007, India.}\\
  {\textit {email: paddy@iucaa.in}}
  }
   
  \date{ }  

\begin{document}
  
  \maketitle
  
  \begin{abstract}
  Principle of Equivalence makes effects of \textit{classical} gravity vanish in local inertial frames. What role does the Principle of Equivalence play as regards \textit{quantum} gravitational effects in the local inertial frames? I address this question here from a specific perspective. 
  At mesoscopic  scales close to, but somewhat larger than, Planck length one could describe quantum spacetime and matter in terms of an effective geometry. 
  The key feature of such an effective quantum geometry is the existence of a \zpl. 
  When we proceed from quantum geometry to quantum matter, the \zpl\ will introduce corrections in the propagator for matter fields in a specific manner. On the other hand, one cannot ignore the self gravity of matter fields at the mesoscopic scales and this will also modify the form of the propagator. Consistency demands that, these two modifications --- coming from two different directions --- are the same. I show that this non-trivial demand  is actually satisfied. Surprisingly, the principle of equivalence, operating at sub-Planck scales, ensures this consistency in a subtle manner.
 \end{abstract}
 
 \section{Quantum gravity in the locally inertial frame}
 
 Principle of Equivalence tells you that, \textit{classical} gravitational effects can be eliminated at the lowest order, around any event $\mathcal{P}$, by choosing a freely falling frame (FFF). If the length scale associated with background curvature\footnote{At any event $\mathcal{P}$, the $L_{curv}$ could be defined in terms of typical curvature components; for example, we can take $L_{curv}^{-2}=\sqrt{R^{abcd}R_{abcd}}$ evaluated at $\mathcal{P}$. If one chooses the FFF at $\mathcal{P}$, then the gravitational effects will come up at distances $x\gtrsim L_{curv}$} at $\mathcal{P}$ is $L_{curv}$, then the gravitational effects arise only at the second order i.e $\mathcal{O}(x^2/L^2_{curv})$ where $x$ is the typical  distance from $\mathcal{P}$ with spacetime remaining (approximately) flat for $x\ll L_{curve}$. But if you start probing very small scales, viz. $x\gtrsim L_P$ (where $L_P$ is the Planck length) you will experience \textit{quantum} gravitational curvature fluctuations even in FFF (`flat spacetime quantum gravity'). This raises the  question: 
 What role does Principle of Equivalence --- which helped us to choose a FFF and eliminate \textit{classical} gravity  --- play as regards \textit{quantum} gravitational effects?  I will discuss this question from a specific perspective in this work.
 
 Let me start with a quantum, spinless particle of mass $m$ and delineate different length scales relevant to its dynamics. 
 I will assume that (i) $m\ll m_{pl}$ where   $m_{pl}$ is the Planck mass and (ii) the quantum field is living in a spacetime which has a large scale GR descrption with a curvature scale $L_{curv}\gg L_P$ where $L_P$ is the Planck length.
The standard QFT associates a length-scale (`Compton wavelength')  
 $\lambda_c \equiv \hbar/mc$ with the quantum field.  When the self-gravity of the quantum particle is introduced into the picture,  another length scale (`Schwarzschild radius'), $\lambda_g \equiv Gm/c^2$ enters the fray. For elementary particles with mass $m\ll m_{pl}$, we have  $\lambda_g=\lambda_c(m/m_{Pl})^2\ll \lambda_c$. On the other hand, Planck length is the geometric mean of the Schwarzschild radius and the Compton wavelength: $L_P=\sqrt{\lambda_c\lambda_g}$. This leads to the ordering of these three lengths as: $\lambda_g < L_P<\lambda_c$. The background geometry introduces one more length-scale $L_{curv}$. In the situations of interest to us here, therefore, we have the ordering:
 \begin{equation}
  \lambda_g < L_P<\lambda_c<L_{curv}
 \end{equation} 
 
 This allows us to disentangle different aspects of dynamics. To start with, standard QFT in flat spacetime will be a good approximation for the field modes with wavelength $\lambda$ if $L_P\ll \lambda\ll L_{curv}$. As we increase $\lambda$ and it becomes comparable to $L_{curve}$ (i.e when $L_P\ll \lambda\approx L_{curv}$), we need the descrption of QFT in CST; the geometry can be treated as classical but it could, for example, produce quanta of the field. (This is an IR regime quantum effect.)
 On the other hand, when we decrease $\lambda$ and approach $\lambda\to L_{P}$ we expect \textit{quantum} gravitational curvature
 effects to come up \textit{even in the FFF}. (This is a UV regime quantum effect.) Further, the fact that $\lambda_g<L_P$ has a curious consequence. At scales comparable to $\lambda_g$ one cannot ignore the self-gravity of the particle and the consequent curvature, even if we started with the assumption the the background curvature is ignorable because $\lambda_c\ll L_{curv}$. This is usually considered irrelevant because when $\lambda\approx\lambda_g$ we are already in the sub-Planckian scales.
 
 To sum up,   Principle of Equivalence allows you to escape from curvature effects (by choosing a FFF) at scales $L_P\ll \lambda \ll L_{curv}$.
 When $\lambda\to L_{curv}$, we cannot ignore \textit{classical} gravitational 
 effects and we need QFT in CST; Principle of Equivalence cannot help at these scales to eliminate classical curvature effects. 
 On the other hand when $\lambda\to L_{P}$ we cannot ignore \textit{quantum} gravitational curvature
 effects.
 The question arises as to whether Principle of Equivalence has any meaningful role to play in this regime. I will show that it does.
 
 \section{Encoding the QG effects at the mesoscopic scales}
 
 While studying  the dynamics, when the modes of the field approach the Planck scales,
  it is useful to distinguish between two regimes, which I will call \textit{microscopic} and \textit{mesoscopic}. 
 The mesoscopic regime interpolates between the  microscopic regime, very close to Planck scale (which requires a full quantum gravitational description) and macroscopic regime, far away from the Planck scale (at which one can use the formalism of quantum field theory in a classical, curved, background spacetime). This regime 
 is close, but not too close,  to the Planck scale so that we can still introduce  some kind of effective geometric description, incorporating quantum gravitational effects to the leading order.  
 
 There are \textit{two} distinct features which come into play in the mesoscopic regime, as we approach the Planck scale. The first, which is well-recognized, is the fact that spacetime close to Planck (and sub-Planck) scales needs to be described very differently from spacetime at macroscopic scales. Much of the work in the area of quantum gravity, indeed,  has something to say about this issue. The second feature --- which has not been  equally emphasized --- concerns the matter sector: \textit{How do you describe matter --- say, an electron --- close to and below Planck scales?} This question is non-trivial because no field --- even classically --- is ever free. All fields possess energy  which curves the spacetime in which it is propagating. It is easy to see that this nonlinearity through self-gravity cannot be ignored  as we approach and cross Planck scale. 
 
 These two features are also conceptually distinct.  The first feature  is related to how the (effective) quantum geometry affects the matter while the second feature is related to how matter at Planck scales modifies the geometry. Nevertheless, consistency demands that we should arrive at the fundamentally same description from either direction. I will show  that this is indeed what happens; both features lead us to an effective quantum (corrected) geometry which exhibits a \zpl\ in the spacetime.  Surprisingly, the  principle of equivalence plays an interesting and subtle role in this description.
 
 \subsection{Three routes to the Propagator}
 
 Consider a scalar field of mass $m$ which is propagating in a space(time) with metric $g_{ik}$ and is treated within the context of  quantum field theory in curved spacetime. I want to work with a descriptor of the dynamics of this field which is robust enough to survive (and be useful) at mesoscopic scales.  The propagator for the field is a good choice for such a description.
 All the physics of the scalar field is contained in the standard Feynman propagator $G_{std}(x_2,x_1)$, or equivalently in the rescaled propagator $\mathcal{G} \equiv m G_{\rm std}$ which will turn out to be simpler to handle algebraically.\footnote{\textit{Notation:} I use the subscript `std' for quantities pertaining to a classical gravitational background, not necessarily flat spacetime; the subscript `QG' gives corresponding quantities with quantum gravitational correction. While dealing with expressions corresponding to a free quantum field in flat spacetime I use the subscript `free'.}  There are three equivalent ways of defining this propagator \textit{without using the notion of a local quantum field operator}. 
 The first definition of the (Euclidean) propagator\footnote{I will work in a Euclidean space(time) for mathematical convenience and will assume that the results in spacetime arise through analytic continuation. This is \textit{not} essential and one could have done everything in the Lorentzian spacetime itself; it just makes life easier.} is: 
 \begin{equation}
\mathcal{G}_{\rm std}(x,y;m)\equiv
mG_{std}(x,y;m^2)=\int_0^\infty m\ ds\ e^{-m^2s}K_{std}(x,y;s)
\label{a14}
 \end{equation} 
 where $K_{\rm std}$ is the standard, zero-mass, Schwinger (heat) kernel given by $K_{\rm std} (x,y;s) \equiv \bk{x}{e^{s\Box_g}}{y}$. Here $\Box_g$ is the Laplacian in the background space(time).
The heat kernel is a purely geometric object, entirely determined by the background  geometry; all the information about the scalar field is contained in the single parameter $m$. It has the structure (in $D=4$):
\begin{equation}
 K_{std}(x,y;s)\propto\frac{e^{-\bar\sigma^2(x,y)/4s}}{s^2}\left[1+ \text{curvature corrections}\right]
\end{equation} 
where $\bar\sigma^2(x,y)$ is the geodesic distance and the curvature corrections, encoded in the Schwinger-Dewitt expansion, will involve powers of $(s/L_{curv}^2)$. The exponential $e^{-m^2s}$ in \eq{a14} suppresses the contributions for $s\gtrsim \lambda_c^2$ in the integral in \eq{a14} and hence when $\lambda_c\ll L_{curv}$, the curvature corrections will be small.

The second definition of the propagator is based on the path integral sum: 
 \begin{equation}
\mathcal{G}_{\rm std}(x_1,x_2;m) = \sum_{\rm paths\ \sigma} \exp -m\sigma (x_1,x_2)
  \label{three}
 \end{equation}
 where $\sigma(x_1,x_2)$ is the length of the path connecting the two events $x_1, x_2$ and the sum is over all paths connecting these two events.
 This path integral can be defined in the lattice and computed --- with suitable measure --- in the limit of zero lattice spacing \cite{pid,tpqft}. The result will agree with that in \eq{a14}.
 
 The third definition is a variant of this, obtained by converting the path integral to an \textit{ordinary} integral. To do this, I will 
 introduce a Dirac delta function into the path integral sum in \eq{three} and use the fact that both  $\ell$ and $\sigma$  are positive definite, to obtain:
 \begin{equation}
\mathcal{G}_{\rm std}(x_1,x_2;m) = \int_{0}^\infty d\ell\ e^{-m\ell} \sum_{\rm paths\ \sigma}\delta_D \left(\ell - \sigma (x_2,x_1)\right)
 \equiv \int_{0}^\infty d\ell\ e^{-m\ell} N_{std}(\ell; x_2,x_1)
  \label{k2}
 \end{equation}
 where we have defined the function $N_{std}(\ell; x_2,x_1)$ to be:
 \begin{equation}
N_{std}(\ell; x_2,x_1) \equiv \sum_{\rm paths\ \sigma}\delta_D \left(\ell - \sigma (x_2,x_1)\right)
  \label{k3}
 \end{equation}
 The last equality in \eq{k2}
 describes the path integral as an \textit{ordinary} integral with a measure $N(\ell)$ which --- according to \eq{k3} --- can be thought of as counting the \textit{effective} number of paths\footnote{Of course, the \textit{actual} number of paths, of a given length connecting any two points in the Euclidean space, is either zero or infinity. But the \textit{effective} number of paths $N(\ell)$, defined as the inverse Laplace transform of $\mathcal{G}$ (see \eq{k2}), will be a finite quantity.} of length $\ell$ connecting the two events $x_1$ and $x_2$. Most of the time I will just write $N(\ell)$ without displaying the dependence on the spacetime coordinates for notational simplicity. 
 
 Before proceeding further, let me illustrate the form of $N(\ell)$ in the case of a free field in flat space. Expressing both $\mathcal{G}_{\rm free}(p,m)=m(p^2+m^2)^{-1}$ and $N_{\rm free}(p,\ell)$ in momentum space, we immediately see that:
 \begin{equation}
  \mathcal{G}_{\rm free}(p^2,m) = m G_{\rm free}(p^2,m^2) = \frac{m}{m^2+p^2} = \int_0^\infty d\ell\ e^{-m \ell} \cos p\ell
   \label{thirteen}
  \end{equation} 
  showing  that $N_{\rm free}(p,\ell)$ in momentum space is given by the simple expression $N_{\rm free}(p, \ell)$  $= \cos(p\ell)$. (The form of $N_{\rm free}(\ell, x_2,x_1)$ in real space can be computed by a Fourier transform; the calculation and the result are given in the Appendix.)
 
 \subsection{Quantum gravity corrections to the Propagator at Mesoscopic scales: Inserting the Planck length}
 
 This description in terms of a propagator, defined by any of the three approaches is totally adequate to handle the matter field, when it is propagating in a given curved spacetime. 
 \textit{None of these definitions use the formalism of a local field theory and its canonical quantisation, notions which may not survive close to Planck scales;} therefore the propagator, defined in any of these three ways, provides a robust construct which we can rely on at mesoscopic scales.
 
 In particular, we can ask:
 What happens to the propagator when we approach the Planck scales? Obviously, the classical geometrical description needs to be modified close to Planck scales in a  manner which is at present unknown. It is, however, possible to capture the most important effects of quantum gravity by introducing a \zpl\ to the spacetime \cite{zplimp}. This is based on the idea that the \textit{dominant} effect of quantum gravity at mesoscopic scales can be captured by assuming\footnote{Such an idea has been introduced and explored extensively in the past literature \cite{zplimp, zplextra} and hence I will not pause to describe it here; I will just accept it as a working hypothesis and proceed further.} that the path length $\sigma^2(x_2,x_1)$ has to be replaced by $\sigma^2(x_2,x_1) \to \sigma^2(x_2,x_1)+ L^2$ where  $L^2$ is of the order of Planck area $L_P^2 \equiv (G\hbar/c^3)$. 
 
  It is easy to see how the introduction of \zpl\ into the  geometry modifies the propagator in \eq{k2}. The existence of the \zpl\ suggests that we change the path length $\ell$ appearing in the amplitude to $(\ell^2 + L^2)^{1/2}$. Therefore the quantum corrected propagator will be given by the last integral in \eq{k2} with this replacement. This leads to the expression for the propagator in an (effective) quantum geometry:
 \begin{equation}
  \mathcal{G}_{\rm QG} (x_1,x_2; m) = \int_0^\infty d\ell \ N_{std}(\ell; x_1,x_2) \exp\left( - m \sqrt{\ell^2+L^2}\right)
  \label{b61}
 \end{equation} 
 The  modification $\ell \to (\ell^2 + L^2)^{1/2}$ ensures that all path lengths are bounded from below by the \zpl.\footnote{One can also obtain the same result by modifying $N_{std}$ to  another expression $N_{QG}$ and leaving the amplitudes the same. But the above interpretation is more intuitive; see Appendix for the connection between the two approaches.} 
 
 We know that the original path integral in \eq{k2} had an equivalent description in terms of the heat kernel through \eq{a14}.  How does the modification in \eq{b61} translate to the relation between the heat kernel and the propagator?  With some elementary algebra, involving Laplace transforms (see Appendix for details), one can show that \eq{a14} is now modified to: 
 \begin{equation}
  \mathcal{G}_{QG}(x,y;m)=\int_0^\infty m\ ds\ e^{-m^2s-L^2/4s}K_{std}(s; x,y)
  \label{a15}
 \end{equation} 
 Recall that  the leading order behaviour of the heat kernel is $K_{\rm std} \sim s^{-2}\exp[-\sigma^2(x,y)/4s]$ where $\sigma^2$ is the geodesic distance between the two events; so the modification in \eq{a15} amounts to the replacement $\sigma^2 \to \sigma^2 +L^2$ to the leading order. That makes perfect sense. 
 
 Again, let me illustrate both \eq{b61} and \eq{a15} --- which are valid in arbitrary curved spacetime --- in the context of a free field in  flat spacetime. Working in the momentum space and using the result $N_{free}(p,\ell)=\cos pl$ in \eq{b61}, we get:
 \begin{equation}
  \mathcal{G}_{\rm QG}(p^2) =  \int_0^\infty d\ell \ e^{-m \sqrt{L^2+\ell^2}} \ \cos(p\ell) = \frac{mL}{\sqrt{p^2+ m^2}} K_1[ L\sqrt{p^2+m^2}]
   \label{fifteen1}
  \end{equation} 
 Similarly, using the expression for momentum space, zero-mass,  kernel in flat space, $K_{std}(s;p)=\exp(-sp^2)$ in \eq{a15} we get:
 \begin{equation}
   \mathcal{G}_{\rm QG}(p^2) = \int_0^\infty ds \ m\ \exp\left[-s(p^2+m^2) - \frac{L^2}{4s}\right] = \frac{mL}{\sqrt{p^2+m^2}} \ K_1[L\sqrt{p^2+m^2}]
  \end{equation} 
which is identical to \eq{fifteen1}. These expressions describe what could be called QG corrections to the propagator in a FFF. 

\subsection{Quantum gravity corrections from another perspective: Discovering the Planck length}
 
 I will now approach the same issue from a different direction.   The  action for a relativistic particle of \textit{inertial} mass $m_i$ gives the factor $\exp(-A/\hbar)$ with $A/\hbar = -m_ic\sigma/\hbar = - \sigma /\lambda_c$ where $\sigma $ is the length of the path and $\lambda_c = \hbar/m_ic$ is the Compton wavelength of the particle.  The Compton wavelength $\lambda_c = \hbar/(m_i c)$ is defined in terms of the \textit{inertial} mass of the particle.  The part of the path integral amplitude $\exp[-(\sigma/\lambda_c)]$ comes from combining special relativity with quantum theory and does not depend on the existence of gravity.  The path integral amplitude  is exponentially suppressed for paths longer than the Compton radius $\lambda_c \equiv \hbar/m_ic$.

 When the self-gravity of the matter field is introduced into the picture,  another length scale, viz. the gravitational Schwarzschild radius $\lambda_g \equiv Gm_g/c^2$ where $m_g$ is the gravitational mass of the particle, comes into play. The self-gravity of a particle of mass $m_g$ will strongly curve the spacetime at length scales comparable to $\lambda_g$. At  length scales comparable to $\lambda_g$, we can no longer think of a `free field' even in  flat spacetime. \textit{In fact, it makes absolutely no sense to sum over paths with $\sigma \lesssim \lambda_g$ in the path integral.}  Just as paths with $\sigma \gtrsim \lambda_c$ are suppressed exponentially by the factor $\exp[-(\sigma/\lambda_c)]$, we should suppress 
 the paths with $\sigma \lesssim \lambda_g$  by another dimensionless factor $F[(\lambda_g/\sigma)]$ which depends on the dimensionless ratio $(\lambda_g/\sigma)$ and rapidly decreases for $\sigma\ll\lambda_g$. This will modify the amplitude for a path of length $\sigma$ from $\exp[-(\sigma/\lambda_c)]$ to $F[(\lambda_g/\sigma)]\exp[-(\sigma/\lambda_c)]$. Writing, $F\equiv\exp -f$ for algebraic convenience, the modified propagator is now given by the path integral sum:
 \begin{equation}
  \mathcal{G} (x,y) \equiv \sum_{\rm paths\ \sigma} \exp\left[ -\frac{\sigma}{\lambda_c} -f[(\lambda_g/\sigma)]\right]= 
 \sum_{\rm paths\ \sigma} \exp-m_i\left[\sigma + \frac{1}{m_i}f[(\lambda_g/\sigma)]\right]
 \label{pid0}
 \end{equation} 
 
 We now have two completely independent ways of defining the propagator at mesoscopic scales. (i) First, 
 starting from the modifications of the quantum geometry and approaching the matter sector we argued that the propagator has to be modified into the form in \eq{b61} or, equivalently, to \eq{a15}. In this approach we introduced the Planck length by hand, through the postulate of \zpl . (ii) Second, starting from matter sector and incorporating the self gravity of a particle of mass $m$  into the path integral propagator, we have arrived at the modification of the propagator in \eq{pid0}. We have not introduced the notion of Planck length explicitly and have only used the two length scales associated with the mass of the particle we are studying.
 
 Consistency demands that these two propagators should be identical, which put \textit{two} nontrivial constraints on expression in \eq{pid0}.
 
 The first constraint allows us to fix the form of the function $F=\exp -f$. Since the result in \eq{b61} has a purely geometrical origin, the \eq{pid0} can reproduce \eq{b61} only if the factor in the square bracket multiplying $m_i$ in \eq{pid0} is just a function of $\sigma$. That is, this factor  cannot depend on the parameters of the scalar field like $m_i, m_g$. This, in turn, is possible only if (i) the Principle of Equivalence holds, allowing us to set $m_i=m_g$ \textit{and} (ii) the function is given by $f[(\lambda_g/\sigma)]\propto (\lambda_g/\sigma)$. The proportionality constant will be of order unity; this is because the paths with lengths
 $\sigma<\lambda_g$ are now suppressed exponentially by the factor $F=\exp -f$ and we expect this suppression to happen for $\sigma\lesssim\lambda_g$. So the proportionality factor can be ignored with the understanding that we now redefine $\lambda_g$ as $\mathcal{O}(1)(Gm/c^2)$. We can thus conclude that  a natural and minimal modification of the path integral sum in \eq{three},  which incorporates the self gravity of a particle of mass $m=m_i=m_g$, will lead to the propagator: 
 \begin{equation}
  \mathcal{G} (x,y) \equiv \sum_{\rm paths\ \sigma} \exp\left[-\frac{\sigma}{\lambda_c}\right]\ \exp\left[-\frac{\lambda_g}{\sigma}\right]
  = \sum_\sigma \exp\left[- m \left(\sigma + \frac{L^2}{\sigma}\right)\right]
  \label{pid}
 \end{equation}   
 where $L= \mathcal{O}(1) L_P$.
  This modification, given by \eq{pid} has \cite{pid} a beautiful symmetry: The amplitude is invariant  under the duality transformation $\sigma \to L^2/\sigma$.
 
 The result  depends on the principle of equivalence in a subtle and interesting way. The Compton wavelength $\lambda_c = \hbar/(m_i c)$ is defined in terms of the \textit{inertial} mass of the particle and gives part of the path integral amplitude $\exp[-(\sigma/\lambda_c)]$, which comes from combining special relativity with quantum theory; this factor does not depend on the existence of gravity. On the other hand, the gravitational radius $\lambda_g \equiv Gm_g/c^2$ is defined in terms of the \textit{gravitational} mass of the particle and leads to the factor $\exp[-(\lambda_g/\sigma)]$. These two factors exist separately in the first equality of \eq{pid}. But they can be expressed as in the second equality of \eq{pid} only because of the assumption $m_i = m_g$! 
  If $m_i \neq m_g$ then we will end up with the argument of the exponential:
 \begin{equation}
  \frac{m_i\sigma}{\hbar c} +\frac{Gm_g}{c^2\sigma} =\frac{1}{\lambda_c}\left[\sigma+\left(\frac{m_g}{m_i}\right)\frac{L_P^2}{\sigma}\right]
 \end{equation} 
 Clearly, there is no universal, geometrical  interpretation for such a factor in the square bracket, occurring in a path integral. The addition of a universal \zpl\ to the spacetime --- which is independent of any parameters of the matter sector --- will \textit{not} be equivalent to the modification of the propagator due to its self-gravity if $m_i \neq m_g$.  \textit{Just as classical gravity admits a purely geometrical description only because $m_i = m_g$, the quantum geometry allows a universal description in terms of \zpl\ only because of $m_i = m_g$.} We now have principle of equivalence operating at Planck scales! So the duality symmetry for $\sigma\to L^2/\sigma$ is closely related to the principle of equivalence.\footnote{This result also tells us why the \textit{exponential} form of the suppression $\exp[-(\lambda_g/\sigma)]$ --- rather than some other functional form --- in \eq{pid}, for path lengths smaller than Schwarzschild radius, is uniquely selected. No other functional form will lead to the geometrical factor $[\sigma + (L^2/\sigma)]$, which is required.} 
 
\textit{The above argument, in a way also ``discovers'' Planck length.} The first equality in \eq{pid} gives two exponential suppression factors, based on two length scales $\lambda_c$ and $\lambda_g$ associated with  the particle. \textit{Both factors depend on the mass of the particle.}. But when combined together, as in the second equality, the Planck length appears (essentially as the geometric mean $L_P=\sqrt{\lambda_c\lambda_g}$) which is independent of the mass of the particle and a universal constant. As a bonus, the duality structure, \textit{with respect to $L_P$}, emerges.\footnote{To be precise we only know that the amplitude is suppressed for path lengths below $\mathcal{O}(1)(Gm/c^2)$; therefore, strictly speaking $L$ and $L_P$ can differ by a factor of order unity. This makes no difference to our analysis and I will not bother to distinguish between $L$ and $L_P$.}
 
 Let me now mention the  second constraint on our construction which is more nontrivial. The path integral sum in \eq{pid} should lead to the same propagator as the one in \eq{b61}. \textit{Remarkably enough, it does!} One can indeed give meaning to the path integral sum in \eq{pid}  by defining it on a lattice and then taking the limit of zero lattice spacing. Such an exercise (see Ref. \cite{pid}) shows that the path integral sum in \eq{pid} \textit{does} lead precisely to the result in \eq{a15}. 
 This result is  non-trivial and could not have been ``guessed''.

 For the sake of completeness, I will mention an alternate way of relating the two directions of approach we have adopted above. To do this, I begin by relating the two propagators $G_{\rm QG}$ and $G_{\rm std}$. It is straightforward to show, again using some Laplace transform tricks, that (see Appendix)
 \begin{align}
  G_{\rm QG}(x,y;m^2)= - \frac{\partial}{\partial m^2} \int_{m^2}^\infty dm_0^2 \ J_0\left[L \sqrt{m_0^2-m^2}\right] G_{\rm std}( x,y;m^2_0)
   \label{five1}
  \end{align}
 This is equivalent to assuming that --- close to Planck scales --- there is an amplitude $\amp{m}{m_0}$  for a system with mass $m_0$ to appear as a system with mass $m$. Such a feature can arise due to quantum fluctuations in the length scales as follows.
If we put $m_0 = \lambda m$ and write $G_{\rm std}$ as a path integral sum, then the above relation can be re-expressed in the form 
\begin{equation}
  G_{\rm QG}(m) = \int_1^\infty d\lambda\ \mathcal{A}(m,\lambda)\, \sum_{\rm paths\ \sigma} e^{-m\lambda \sigma}
  \label{z3}
  \end{equation}
  with 
  \begin{equation}
  \mathcal{A}(m,\lambda) = - \frac{\lambda(Lm)}{\sqrt{\lambda^2 -1}}\ J_1\left[m L\sqrt{\lambda^2-1}\,\right] 
  \label{z4}
  \end{equation}
  for $\lambda>1$. (There is a Dirac delta function contribution at $\lambda=1$ which I have not displayed; see Appendix) This suggests the following interpretation: The presence of a mass $m$ in the space(time) induces fluctuations in the length scales changing $\sigma \to \lambda \sigma$ with an amplitude $\mathcal{A}(m, \lambda)$. The correct propagator $G_{\rm QG}(m)$ has to be obtained by integrating over these fluctuations as well as the sum over paths along the lines of \eq{z3}. 
These results tell us that as we approach Planck scales, fluctuations of quantum geometry and quantum fluctuations of matter merge with each other and acquire a unified description in terms of the \zpl. 
 
 \section{Conclusions}
 
 Consider a region of spacetime in which the curvature length scale $L_{\rm curv}$ is much larger than Planck length: $L_{\rm curv} \gg L_P$. Concentrate on the modes of  a quantum field which probe the several orders of magnitude between $L_P$ and $L_{\rm curv}$. Let us start with modes which are far away from either extremities: $L_P \ll \lambda \ll L_{\rm curv}$, and study them in the freely falling frame (FFF) around an event $\mathcal{P}$ in this spacetime region. The \textit{classical} effects due to spacetime curvature will be absent to order $\mathcal{O} (\lambda^2/L_{\rm curv}^2)$. The Principle of Equivalence, which allows the choice of FFF around any even $\mathcal{P}$, has eliminated classical gravity.
 
 Let us now start decreasing $\lambda$. Since we are in FFF, no classical gravitational effects due to curvature can arise and the approximation of a flat spacetime becomes more and more accurate as $\lambda$ becomes progressively smaller compared to $L_{\rm curv}$. But when we start approaching Planck length (i.e., when $\lambda \approx C L_P$ where $C$, say, is about $10^2$) \textit{quantum} gravitational effects will start appearing. However, we are still immune to \textit{classical} gravitational effects because we are working in flat spacetime to a high order of accuracy. This is the regime of  \textit{flat spacetime quantum gravity} around any event $\mathcal{P}$. 
 
 There is an alternative way of understanding this effect, again as a consequence of Principle of Equivalence. One formulation of Principle of Equivalence will be to postulate that laws of classical special relativity will remain valid in a FFF around any event $\mathcal{P}$. 
 But a classical flat spacetime will harbor quantum gravitational fluctuations, just as a classical electromagnetic vacuum will harbor quantum electrodynamical fluctuations.  The Principle of Equivalence tells us
  that the quantum gravitational effects in FFF will be identical to the quantum gravitational effects in a (globally) flat spacetime. The effect of background curvature can be ignored to the order $\mathcal{O}(L_P^2/L_{\rm curv}^2)$. Of course, if you want to study situations in which $L_{\rm curv} \approx L_P$, you need the full machinery of quantum gravity; but when $L_{\rm curv} \gg L_P$ we can still meaningfully talk about quantum gravitational effects adding corrections to standard QFT in the mesoscopic regime with $\lambda$ close --- but not too close --- to $L_P$. The delineation of this flat spacetime quantum gravity regime is one of the important conceptual results of this paper.

 The next step is to ask how quantum gravity affects standard flat spacetime QFT in the mesoscopic regime. The most important feature is the inclusion of a \zpl\ into the propagator and through that into the dynamics of the quantum field. To do this, I introduced another  useful concept, viz., that of number of effective quantum paths $N_{\rm std} (\ell)$. This is defined by \eq{k2} [or \eq{3mar} in the Appendix] as the Laplace transform of the rescaled  propagator $\mathcal{G}_{\rm std} \equiv m G_{\rm std}$, in a given classical curved background, with respect to the mass $m$. These defining equations have a simple physical interpretation: The equation
 \begin{equation}
 \mathcal{G}(x_2,x_1) \equiv m G(x_2,x_1) \equiv \int_0^\infty d\ell\ N(\ell; x_2,x_1)\, e^{-m\ell}
 \end{equation} 
tells us that:
{\small{
\begin{equation*}
\left[
\begin{array}{c}
  \text{Amplitude to propagate}\\
  \text{from}\ x_1\ \text{to}\ x_2
\end{array}
\right]
= \int_0^\infty d\ell
\left[
\begin{array}{c}
  \text{Effective number}\\
  \text{of paths of length}\ \ell
\end{array}
\right]
\times
\left[
\begin{array}{c}
  \text{Amplitude associated}\\
  \text{with a path of length}\ \ell
\end{array}
\right]
\end{equation*}
}}
 
 Incorporating the \zpl\ involves changing the amplitude for a given path of length $\ell$ from $\exp(-m\ell)$ to $\exp(-m\sqrt{\ell^2+L^2})$. This leads to the quantum corrected propagator incorporating the \zpl\ in a given background curved spacetime:
  \begin{equation}
  \mathcal{G}_{\rm QG} (x_1,x_2; m) = \int_0^\infty d\ell \ N_{std}(\ell; x_1,x_2) \exp\left( - m \sqrt{\ell^2+L^2}\right)
  \label{b61new}
 \end{equation}
 That is,
 {\small{
 \begin{equation*}
 \left[
 \begin{array}{c}
  \text{Propagation amplitude}\\
  \text{incorporating}\\
  \text{QG corrections}
 \end{array}
 \right]
 = \int_0^\infty d\ell
\left[
 \begin{array}{c}
 {}\\
  \text{Effective number}\\
  \text{of paths of length}\ \ell \\
  {}
 \end{array}
\right]
\times
\left[
\begin{array}{c}
 \text{Amplitude associated}\\
 \text{with a path of length}\ \ell\\
 \text{incorporating}\\
\text{\zpl}
\end{array}
\right]
 \end{equation*}
 }}%
Alternative representations for the same result are given by \eq{a15} and in terms of $N_{\rm QG}$ introduced in \eq{nqgmu} of the Appendix. In this approach to quantum gravitational corrections to the propagator, the \zpl\ is introduced into the spacetime and it modifies the propagator. 
 
 As described in the earlier sections, Principle of Equivalence also allows us to start from the scalar field of mass $m$ and ``discover'' the Planck length. We start with two natural length scales associated with mass $m$: The Compton radius $\lambda_c \equiv \hbar/m_ic$ (where $m_i$ is the \textit{inertial} mass of the particle) and the Schwarzschild radius $\lambda_g \equiv Gm_g/c^2$ (where $m_g$ is the gravitational mass of the particle). The standard path integral for a quantum field in curved spacetime associates an amplitude $\exp(-\ell/\lambda_c)$ with a path of length $\ell$; this shows that quantum field theoretic correlations are suppressed at $\ell \gg \lambda_c$. On the other hand, it makes no sense to sum over paths where $\ell < \mathcal{O}(1) \lambda_g$. This suggests introducing an extra factor $\exp(-\lambda_c/\ell)$ into the path integral amplitude.  This modifies the form of the propagator to 
   \begin{equation}
   \mathcal{G} (x,y) \equiv \sum_{\rm paths\ \sigma} \exp\left[-\frac{\sigma}{\lambda_c}\right]\ \exp\left[-\frac{\lambda_g}{\sigma}\right]
 \end{equation}
 That is, 
 {\small{
 \begin{equation*}
 \left[
  \begin{array}{c}
   \text{Propagation amplitude}\\
   \text{incorporating}\\
   \text{QG corrections}
  \end{array}
  \right]
  =\mathlarger{\mathlarger{\sum}}_{\rm paths}
\left[
\begin{array}{c}
 \text{Standard amplitude with}\\
 \text{exponential suppression}\\
 \text{for}\  \sigma > \lambda_c; \text{depends on}\\
 \text{inertial mass}\ m_i
\end{array}
\right]
\times
\left[
\begin{array}{c}
 \text{Amplitude with}\\
 \text{suppression for}\\
  \sigma < \lambda_g; \text{depends on}\\
 \text{gravitational mass}\ m_g
\end{array}
\right]
 \end{equation*}
 }}%
 The Principle of Equivalence, stated as $m_i = m_g$, allows us to combine these two factors and write the quantum gravity corrected propagator in a purely geometrical form: 
 \begin{equation}
  \mathcal{G}_{QG}(x,y;m) 
  = \sum_\sigma \exp\left[- m \left(\sigma + \frac{L^2}{\sigma}\right)\right]
  \end{equation}
 We have ``discovered'' the Planck length as the geometric mean of Compton radius and Schwarzschild radius, because the principle of equivalence ($m_i = m_g$) allows the mass  to disappear in the equation  $L_P = \sqrt{\lambda_g\lambda_c}$. 
 
 The key point to note is that, the propagators that incorporate the  quantum gravitational corrections, obtained by these two different routes are identical! (I have, of course, anticipated this result by using the same symbol $\mathcal{G}_{\rm QG}$ for both propagators in the relevant equations.)  Once again, Principle of Equivalence plays a subtle role in leading to  these results  at  mesoscopic scales.
 
 \section*{Acknowledgement}

I thank Sumanta Chakraborty and Dawood Kothawala comments on an earlier draft.  My research  is partially supported by the J.C.Bose Fellowship of Department of Science and Technology, Government of India.

\appendix

\section*{Appendix:  Calculational details}

This Appendix gives the details of some of the calculations as well as some alternative derivations and extensions.

\section{Relation between $G_{QG}$ and $G_{std}$}

The inclusion of \zpl\ modifies  standard (rescaled) propagator 
\begin{equation}
  \mathcal{G}_{std}(x,y;m)\equiv mG_{std}=\int_0^\infty m\ ds\ e^{-m^2s}K_{std}(s; x,y)
  \label{a15new1}
 \end{equation}
to the form $\mathcal{G}_{QG}$ which incorporates the quantum corrections:
\begin{equation}
  \mathcal{G}_{QG}(x,y;m)=\int_0^\infty m\ ds\ e^{-m^2s-L^2/4s}K_{std}(s; x,y)
  \label{a15new}
 \end{equation} 
 There are two ways of understanding this result. The simple, intuitive way is to recall that the leading order behaviour of the heat kernel is $K_{\rm std} \sim s^{-2} \exp[-\sigma^2(x,y)/4s]$ where $\sigma^2$ is the geodesic distance between the two events; so the modification in \eq{a15new} amounts to the replacement $\sigma^2 \to \sigma^2 +L^2$ to the leading order. This gives the leading QG corrections to the propagator at mesoscopic scales. The corrections due to background curvature, captured in the Schwinger-DeWitt coefficients are irrelevant at the mesoscopic scales $\lambda$ with
 $L_P\lesssim\lambda\ll L_{curv}$; this is what I called the flat spacetime quantum gravity regime. 
 
 More rigorously, one can arrive at \eq{a15new} from the principle of path integral duality. This principle postulates \cite{pid} that the effect of \zpl\ is to modify the path integral to the form:
\begin{equation}
  \mathcal{G}_{QG}(x,y;m) 
  = \sum_\sigma \exp\left[- m \left(\sigma + \frac{L^2}{\sigma}\right)\right]
  =\int_0^\infty m\ ds\ e^{-m^2s-L^2/4s}K_{std}(s; x,y)
  \label{pid1}
 \end{equation} 
The path integral  sum can be computed by lattice regularization techniques \cite{pid} and will lead to the second equality. 

I will briefly outline how straight forward algebra allows one to relate $G_{QG}$ and $G_{std}$.
We start with a standard integral involving Bessel function:
\begin{equation}
   \int_0^\infty 2K dK \ J_0(KL) e^{-sK^2} = \frac{1}{2} \exp\left( -\frac{L^2}{4s}\right)
   \label{ythirty}
  \end{equation} 
and obtain from this the result  
 \begin{align}
 \int_a^\infty dt \ e^{-pt} J_0\left[ 2 \sqrt{b(t-a)}\right] = \frac{1}{p} e^{-ap - (b/p)}
   \label{1mar} 
  \end{align} 
  which can be verified by setting $b(t-a) = x^2 $ and using \eq{ythirty}.
  Differentiating both sides of \eq{1mar} with respect to $a$ we get 
    \begin{equation}
 e^{-ap - (b/p)} =  - \frac{\partial}{\partial a} \int_a^\infty dt\ e^{-pt} J_0\left[2 \sqrt{b(t-a)}\right]
   \label{two}
  \end{equation} 
  The limits of integration  in the right hand side can be extended from 0 to $\infty$ by introducing a factor $\theta[t-a]$ in the integrand.
    Moving  the differential operator $\partial/\partial a$ inside the integral, one will then obtain one term containing $\theta J_1$ and another term of the form $J_0 \delta$ giving rise to $e^{-ap}$. 
    It turns out, however, more convenient \textit{not} to do this and instead use the expression in \eq{two} as it is in the computations. The differentiation is best carried out towards the end, when required. I will now set $a=m^2$ and $b=L^2/4$ in \eq{two} to obtain: 
  \begin{align}
  e^{-m^2s - (L^2/4s)} &=  - \frac{\partial}{\partial m^2} \int_{m^2}^\infty dt\ e^{-st} J_0\left[L \sqrt{t-m^2}\right]\nonumber\\
  &=- \frac{\partial}{\partial m^2}\int_{m^2}^\infty d m_0^2\ e^{-m_0^2s} J_0\left[L \sqrt{m_0^2-m^2}\right]
   \label{three1}
  \end{align} 
  where, in the second step, I have put $t=m_0^2$. I insert this expansion in the definition of quantum gravitational propagator, given by
  \begin{equation}
   G_{\rm QG} (m^2) = \int_0^\infty ds\ e^{-m^2s - (L^2/4s)} K_0(s)
   \label{four}
  \end{equation} 
  where $K_0(s) = \bk{x}{e^{s\Box}}{y}$ is the zero-mass heat kernel in an arbitrary curved space(time)\footnote{I have suppressed the dependence of $K_0$ and $G_{\rm QG}$ on the coordinates $x,y$ for notational simplicity. If the metric is independent of some of the coordinates, the same relation can be used in momentum space as well because the integrals for Fourier transform with respect to these coordinates just flow through the expressions in both sides.}  with $\Box \equiv \Box_g$ being the Laplacian corresponding to the curved space metric $g_{ab}$.
  Using \eq{three1} in \eq{four} I obtain:
  \begin{align}
  G_{\rm QG}(m^2) &= - \int_0^\infty ds \ K_0(s) \frac{\partial}{\partial m^2} \int_{m^2}^\infty dm_0^2 \ e^{-m_0^2 s} J_0\left[L \sqrt{m_0^2-m^2}\right]\nonumber\\
  &= -\frac{\partial}{\partial m^2} \int_{m^2}^\infty dm_0^2 \ J_0\left[L \sqrt{m_0^2-m^2}\right] \int_0^\infty ds \ K_0(s) e^{-m^2_0 s}\nonumber\\
  &= - \frac{\partial}{\partial m^2} \int_{m^2}^\infty dm_0^2 \ J_0\left[L \sqrt{m_0^2-m^2}\right] G_{\rm std} (m_0)
   \label{five}
  \end{align} 
  In arriving at the last equality I have used the fact that the standard QFT propagator (without quantum corrections) corresponding to a mass $m_0$  in this space is given by the integral
  \begin{equation}
   G_{\rm std}(m_0) = \int_0^\infty ds \ K_0(s) e^{-m_0^2 s}
   \label{gstdm}
  \end{equation}
  Equation (\ref{five})  relates the quantum corrected propagator for mass $m$ to the standard QFT propagator for mass $m_0$ in an arbitrary Euclidean space(time). Whenever the latter is known, the former can be computed.  
  
 Let us verify this result for the flat space(time) in which $G_{\rm free}$ in momentum space is given by 
  \begin{equation}
   G_{\rm free}(p^2,m_0^2) = \int_0^\infty d\mu\ e^{-\mu(p^2 + m_0^2)}
   \label{six}
  \end{equation} 
  Using this expression in \eq{five}, changing variable to  $x^2 \equiv m_0^2 -m^2$ and carrying out the integrals, we find that
  \begin{align}
    G_{\rm QG}(m^2) & = -  \frac{\partial}{\partial m^2} \int_0^\infty 2 x \ dx\ J_0[Lx] \int_0^\infty d\mu\ e^{-\mu p^2} \, e^{-\mu(m^2 + x^2)}\nonumber\\
    & = -  \frac{\partial}{\partial m^2} \int_0^\infty d\mu\ e^{-\mu(p^2 + m^2)} \int_0^\infty 2 x \ dx\ J_0[Lx]e^{-\mu x^2} \nonumber\\
   & =  \int_0^\infty d\mu\ e^{-\mu(p^2 + m^2)- (L^2/4\mu)}
   \label{seven}
  \end{align} 
  where, to obtain the last equality, I have used the  identity in \eq{ythirty}.
  Clearly, \eq{seven} gives the correct quantum gravity corrected propagator in flat space(time).

  \section{Euclidean Path Measure}

  Let us start with the definition of path measure for  $\mathcal{G}\equiv mG$ through the equation
   \begin{equation} 
  \mathcal{G} \equiv m G \equiv \int_0^\infty d\ell\ N(\ell)\, e^{-m\ell}
   \label{3mar} 
  \end{equation}
  Very often we will work with $\mathcal{G}$ rather than $G$. In Fourier space, $\mathcal{G}$ has the dimensions of length thereby making $N(\ell)$ in \eq{3mar} dimensionless in Fourier space. On Fourier transforming the corresponding path measure in real space acquires the dimension of $L^{-D}$, which allows it to be interpreted as a spacetime density. Of course, it is assumed that  $N$ is independent of $m$; that is,  $N(\ell)$ is treated as the inverse Laplace transform of the rescaled propagator $\mathcal{G}$ from the variable $m$  to variable $\ell$. 
  
  This definition  can be illustrated in the standard free field case taking both the propagator and path measure in the Fourier space. In that case one can easily verify that
  \begin{equation} 
 \mathcal{G}_{\rm free} (p^2,m) \equiv m G_{\rm free} (p^2,m) = \frac{m}{m^2+p^2} = \int_0^\infty d\ell\ N_{free}(\ell) \, e^{-m\ell}
   \label{6mar} 
  \end{equation} 
holds  with  the following choice for $N(\ell)$
  \begin{equation}
  N_{free}(\ell) = \cos p\ell
  \end{equation}
  which is indeed the inverse Laplace transform. (One can also satisfy \eq{6mar}, treated purely as an integral relation by the choice $N(\ell)=e^{-(p^2/m)\ell}$; but this is not acceptable since we want $N(\ell)$ to be independent of $m$. This is why I define $N$ as the inverse Laplace transform of $\mathcal{G}$.)
  
  One can also express the quantum corrected propagator $G_{\rm QG}$ to the path measure $N_{\rm std} (\ell)$  in an arbitrary curved space(time).\footnote{Recall the notation: I use the subscript `std' for quantities pertaining to a classical gravitational background, not necessarily flat spacetime; the subscript `QG' gives corresponding quantities with quantum gravitational correction. While dealing with expressions corresponding to a free quantum field in flat spacetime I use the subscript `free'.} 
  We begin by rewriting \eq{five}  in terms of $\mathcal{G}_{\rm std}$ in the integrand, getting
  \begin{equation}
 G_{\rm QG}(m^2) = - \frac{\partial}{\partial m^2} \int_{m^2}^\infty dm_0^2\ \frac{J_0}{m_0} \, \mathcal{G}_{\rm std}(m_0)
   \label{eight}
  \end{equation} 
  Expressing $\mathcal{G}_{\rm std}$ in terms of $N_{\rm std}$ using \eq{6mar} and multiplying both sides of \eq{eight} by $m$ we get the result
  \begin{align}
  \mathcal{G}_{\rm QG} &= - m \frac{\partial}{\partial m^2} \int_{m^2}^\infty 2 dm_0\ \int_0^\infty d\ell \ N_{\rm std}(\ell)\,  e^{-m_0\ell} J_0 \nonumber\\ 
  & = - \frac{\partial}{\partial m} \int_0^\infty d\ell \ N_{\rm std}\int_{m^2}^\infty  dm_0\ e^{-m_0 \ell} J_0 \left[ L\sqrt{m_0^2 - m^2}\right] \nonumber\\
  &=  - \frac{\partial}{\partial m}\int_0^\infty d\ell \ N_{\rm std}(\ell)\int_1^\infty dx\ m e^{-mx\ell} J_0\left[ mL \sqrt{x^2 -1}\right]
   \label{nine}
  \end{align} 
  The integral can be evaluated using the identity 
  \begin{equation}
  \int_1^\infty dx\ e^{-\alpha x} \, J_0(\beta\sqrt{x^2 -1}) = \frac{e^{-\sqrt{\alpha^2 + \beta^2}}}{\sqrt{\alpha^2+\beta^2}}
   \label{ten}
  \end{equation} 
  to give the rather nice result
  \begin{equation}
  \mathcal{G}_{\rm QG} (x_1,x_2; m) = \int_0^\infty d\ell \ N_{\rm std}(x_1,x_2; \ell) \exp\left( - m \sqrt{\ell^2+L^2}\right)
   \label{eleven}
  \end{equation} 
  This result  tells you that one can interpret the quantum correction involving the zero point length as a simple replacement:  $\ell \to \sqrt{\ell^2 + L^2}$ \textit{without} changing the path measure at  all! This is a viable (though not unique) interpretation. 
   
 An alternative interpretation is to keep the path integral amplitude to be the same (i.e as $\exp -[m\sigma(x,x')]$) but introduce the quantum gravity corrections on the path measure changing $N_{std}$ to $N_{QG}$. To do this, we will use the relation in \eq{3mar} between path measure and the propagator.
  If we use \eq{3mar} with $N (\ell) = N_{\rm std} (\ell)$ we get the standard QFT result $ \mathcal{G}_{\rm std}(m) = m G_{\rm std} (m^2)$; on the other hand, if we use \eq{3mar} with a suitable $N (\ell) = N_{\rm QG} (\ell)$ we should get the quantum corrected propagator $ \mathcal{G}_{\rm QG}(m)= m G_{\rm QG} (m^2)$.  (Here, as everywhere else, I do not explicitly display the space(time) dependencies; to be precise, $\mathcal{G}(m)=\mathcal{G}(x_1,x_2;m)$ and $N (\ell)=N (x_1,x_2;\ell)$.) 
 It is easy to determine  $N_{\rm QG} (\ell)$ by changing  the integration variable in \eq{eleven} from $\ell$ to $\mu$  through $\mu^2 = \ell^2 + L^2$ and rewrite \eq{eleven} in the form: 
  \begin{equation}
  \mathcal{G}_{\rm QG} (x_1,x_2; m)= \int_0^\infty \frac{\mu \, d\mu}{\sqrt{\mu^2-L^2}} \, \theta(\mu - L) \, N_{\rm std} (x_1,x_2; \ell = \sqrt{\mu^2 - L^2} ) \, e^{-m\mu}
   \label{twelve}
  \end{equation} 
  In this form we keep the path integral factor to the standard one $\exp-m\ell$ but change the path measure.
  The quantum corrected path measure $N_{\rm QG}(\mu)$ is related to the standard  QFT path measure $N_{\rm std}(\ell)$ by the simple relation 
  \begin{equation}
   N_{\rm QG}(x_1,x_2; \mu) = \frac{\mu}{\sqrt{\mu^2 - L^2}}\, N_{\rm std}\left[x_1,x_2; \ell = \sqrt{\mu^2-L^2}\right] \theta(\mu - L)
   \label{nqgmu}
  \end{equation} 
In this interpretation,  the path measure for lengths $\mu$ less than the zero point length $L$ are irrelevant to physics and does not contribute. For $\mu > L$, a simple rescaling takes care of the change from $N_{\rm std} $ to $N_{\rm QG}$. It should be stressed that the whole interpretation  depends on $N_{\rm QG}$ being a purely geometrical construct that is \textit{independent} of the mass $m$ of the field, which is clearly seen in the above expression. 
  
  In flat space(time) we can easily verify this result. It is convenient to use the momentum space expressions for the propagators as well as for $N_{\rm free}$ and $N_{\rm QG}$ for this purpose. In flat space(time) we have the result
    (in momentum space)  given by the simple expression $N_{\rm free}(p, \ell) = \cos(p\ell)$. Therefore, \eq{nqgmu} gives the corresponding $N_{\rm QG}(p,\ell)$ to be:
  \begin{equation}
  N_{\rm QG}(p, \ell) = \theta(\ell - L) \frac{\ell \cos p\sqrt{\ell^2- L^2}}{\sqrt{\ell^2-L^2}}
   \label{fourteen}
  \end{equation} 
  So $\mathcal{G}_{\rm QG}(p)$ is given by the integrals in either \eq{eleven} or \eq{twelve}. Using \eq{eleven} we get:
  \begin{equation}
  \mathcal{G}_{\rm QG} = m\, G_{\rm QG}(p, L) = \int_0^\infty d\nu \ e^{-m \sqrt{L^2+\nu^2}} \ \cos(p\nu) = \frac{mL}{\sqrt{p^2+ m^2}} K_1\left[ L(p^2+m^2)^{1/2}\right]
   \label{fifteen}
  \end{equation} 
  which is, of course, the standard result. To arrive at the final result we have used the cosine transform:
  \begin{equation}
   \int_0^\infty dx (\cos bx)\, e^{-\beta\sqrt{\gamma^2 + x^2}} = \frac{\beta\gamma}{\sqrt{\beta^2 + b^2}} \, K_1 \left[ \gamma \sqrt{\beta^2 + b^2}\right]
  \end{equation} 
  
 Finally, let us compute $N_{\rm std}(\ell)$  in real space by a Fourier transform of \eq{fourteen}.  This will lead to the integral:
    \begin{equation}
     N_{\rm QG}(\ell,x) = \frac{\ell \, \theta(\ell -L)}{\sqrt{\ell^2 - L^2}} \, \int \frac{d^D p}{(2\pi)^D} \, e^{ip \cdot x} \ \cos p\sqrt{\ell^2 - L^2}
    \end{equation} 
 To evaluate this expression we need a standard result. If 
  \begin{equation}
 F(k) = \int d^D\bm{x}\ f(|\bm{x}|) \, e^{-i\bm{k\cdot x}}
  \end{equation}
  then we can write
  \begin{equation}
k^{\frac{D-2}{2}} F(k) = (2\pi)^{D/2} \int_0^\infty J_{\frac{D-2}{2}} (kr) \, r^{\frac{D-2}{2}}\ f(r)\, r \, dr
\label{foursevena}
  \end{equation}
   Using \eq{foursevena} and the  cosine transform  we can compute this integral and obtain 
    \begin{equation} 
  I = \int\frac{d^Dp}{(2\pi)^D}\ e^{ip\cdot x} \cos pR = \frac{\theta(R^2 - x^2)}{\pi^{(D-1)/2}} \, \frac{1}{\Gamma\left( - \frac{D-1}{2}\right)} \, \frac{R}{(R^2-x^2)^{(D+1)/2}}
   \label{11mar}  
    \end{equation} 
  This leads to the result
  \begin{equation}
  N_{\rm QG}(\ell,x) = C(D)\ \theta \left[ \ell^2 -(x^2 + L^2)\right] \, \frac{\ell}{ \left[ \ell^2 -(x^2 + L^2)\right]  ^{(D+1)/2}}
   \label{12mar}  
  \end{equation} 
  where
  \begin{equation} 
  C(D) = \frac{1}{\pi^{(D-1)/2}}\, \frac{1}{\Gamma\left( - \frac{D-1}{2}\right)}
   \label{13mar}  
  \end{equation} 
  In $D=4$, this leads to the result
  \begin{equation}
  N_{\rm QG}^{D=4}(\ell,x) = \frac{3}{4\pi^2}\ \theta\left[ \ell^2 - (x^2 + L^2)\right] \, \frac{\ell}{\left[ \ell^2 - (x^2 + L^2)\right]^{5/2}}
   \label{14mar}  
  \end{equation} 
  The result without zero point length $N_{\rm std}$ can, of course, be obtained by putting $L =0$ in these expressions. Note that only paths which contribute are those with a length $\ell^2 > x^2 + L^2$. The singularity structure at $\ell^2 = x^2 + L^2$ should be handled by differentiating the expressions with respect to $L^2$ twice.

 \end{document}